\documentclass[english]{article}
\usepackage{graphicx,psfrag,color}
\usepackage{amsmath} 
\usepackage{amssymb} 
\usepackage{bm} 
\def\bc{\begin{center}}
\def\ec{\end{center}}

\def\beq{\begin{equation}}
\def\eeq{\end{equation}}

\def\bc{{\bf c}}

\def\cb{\color{blue}}

\begin{document}

\title{Optical Transmission of 2D Material with Quantum Anomalous Hall Effect}
\author{Nathan Pravda$^{1}$, Oleg L. Berman$^{1,2}$, Klaus Ziegler$^{2,3}$\\
$^{1}$The Graduate School and University Center, \\
The City University of New York, New York, NY 10016, USA \\
$^{2}$Physics Department, New York City College of Technology, \\
The City University of New York, Brooklyn, NY 11201, USA \\
$^{3}$Institut für Physik, Universität Augsburg, 
86159 Augsburg, Germany \\
}

\maketitle

\begin{abstract}
We study the optical properties of gapped two-dimensional materials which are subject to the 
quantum anomalous Hall effect. At sufficiently low temperatures the transmission, reflection and 
absorption coefficients are found to have a universal behavior that depends only on the ratio of 
the photonic energy and the gap energy. There is a singular behavior with total reflection
when these energies are equal. In the limit of a vanishing gap we recover results for graphene,
where the optical coefficients depend only on the fine-structure constant. The observed optical 
properties provide an accurate measurement of the bandgap.
\end{abstract}
\vspace{0.1cm} 


\section{Introduction}

The optical response of two-dimensional (2D) materials has attracted significant attention in semiconductor physics due to their unconventional 
electronic structure and transport properties. While graphene and related hexagonal systems have been extensively studied, materials 
exhibiting topologically nontrivial transport, particularly the quantum anomalous Hall (QAH) effect, remain an active area of investigation.

The QAH effect is characterized by a finite transverse (Hall) conductivity in the absence of an external magnetic field, arising from intrinsic
time-reversal symmetry breaking. It is closely related to the quantum Hall effect, in which the Hall conductivity is quantized in integer
multiples of $e^2/h$ under an applied magnetic field~\cite{Chang_Liu_Macdonald}. 
In contrast, the QAH effect emerges from the band topology of the material.
Experimentally, QAH phases have been realized in magnetically doped topological insulators, where ferromagnetism is induced via 
transition-metal doping 
(e.g., Cr-, V-, or Mn-doped (Bi,Sb)$_2$Te$_3$) \cite{Chang_Liu_Macdonald,Chang_2,Dyck,Hor,Wang}
and in intrinsic magnetic topological insulators such as Mn-doped (Bi,Sb)$_2$Te$_4$. 
In addition, moiré-engineered systems provide a versatile platform for realizing QAH states. Twisted bilayer graphene 
and transition metal dichalcogenide heterobilayers (e.g., MoTe$_2$/WSe$_2$)~\cite{Sharpe,Li}
can host flat bands with nonzero Chern number, enabling 
QAH phases at fractional fillings; for example, a Chern number $C = 2$ state has been predicted at 2/3 filling in twisted bilayer graphene.

Recent theoretical work has further expanded the class of candidate systems, suggesting that QAH behavior may also arise in ferromagnetic 
metals and even in nonmagnetic systems such as bismuth monolayers with large Chern numbers~\cite{Wan,Zhang}. 
This broadens the range of materials 
available for applications, including resistance metrology, low-power electronics, and spintronic devices~\cite{Wang}. 
Understanding the optical response 
of such systems is therefore of both fundamental and practical interest.

In this work, we investigate the transmission, reflection, and absorption of electromagnetic radiation in a 2D system exhibiting the QAH effect. 
We adopt a two-band model with band gap $2m$, following Ref.~\cite{Nualpijit_Sinner_Ziegler},
and compute the optical response using Maxwell’s 
equations with appropriate boundary conditions. The material is characterized by a conductivity tensor containing both longitudinal and 
Hall components, with the Hall conductivity $\sigma_H$ encoding the QAH response.

The analysis naturally separates into two frequency regimes defined by $\zeta = \hbar \omega / 2m$. 
For $\zeta < 1$, interband transitions are suppressed and the longitudinal conductivity vanishes, leaving 
a purely antisymmetric conductivity tensor with nonzero Hall components. For $\zeta \geq 1$, interband transitions 
lead to a finite longitudinal conductivity, and both diagonal and off-diagonal components contribute. This results 
in a discontinuity in the transmission and reflection coefficients at $\zeta = 1$, along with the onset of absorption.
For comparison, we also consider a system with purely longitudinal conductivity and no QAH effect, described by a 
diagonal conductivity tensor with equal components. In this case, the same threshold behavior at $\zeta = 1$ 
is observed, but without Hall contributions.

The paper is organized as follows. In Sect. \ref{sect:e-field} we present the Maxwell equation for a 2D sheet
in a oscillating electric field. The quantum transport in gapped 2D materials is discussed in 
Sect. \ref{sect:q-transport} and its consequences for the transmission, reflection and absorption
are analyzed in Sect. \ref{sect:o-properties}. This is followed in Sec. \ref{sect:discussion} by a discussion
of the results.

\section{Electric field} 
\label{sect:e-field}

The electric field is subject to the Maxwell equations, in which the 2D material acts as a boundary
condition in $z$ direction at $z=0$, while the field is homogeneous in $x$ and $y$ direction.
We distinguish three fields, namely the incident field $\bm{E_i} =\bm{\tilde{E_i}}e^{iqz+i\omega t}$,
the transmitted field $\bm{E_t} = \bm{\tilde{E_t}}e^{iqz+i\omega t}$ and the reflected field
$\bm{E_r} =\bm{\tilde{E_r}}e^{-iqz+i\omega t}$, where we assume that these fields vary only in the $z$-direction
periodically with wave number $q$.
Since the electric field is continuous at the surface, the boundary condition at $z=0$ reads
\begin{equation} 
\label{bc1}
\bm{\tilde{E_i}} + \bm{\tilde{E_r}} = \bm{\tilde{E_t}}
.
\end{equation} 
The goal is to find a solution of Maxwell's equations with this boundary condition. The source of the field
inside the 2D sheet is given by a current density $\bm j(z,t)=\bm j \delta (z)e^{i\omega t}$, such that
the Maxwell equation becomes
\begin{equation}
\label{maxwell1}
\frac{\partial^2\bm E}{\partial z^2} = -\frac{\omega^2}{c^2}\bm E 
- i\omega\mu_0\delta (z)\bm j
\end{equation}
with the relative dielectric constant $\epsilon$ and the vacuum dielectric constant $\epsilon_0$.
Integrating this equation and using in the continuity equation Eq. (\ref{bc1})
yields~\cite{Nualpijit_Sinner_Ziegler}
\begin{equation} 
\label{maxwell1a}
-iq\bm{\tilde{E_t}} + iq(\bm{\tilde{E_i}} -\bm{\tilde{E_r}}) = i\omega\mu_0\bm j
.
\end{equation} 
Since the current density is the response to the electric field inside the sheet, we
employ Ohm's law $\bm j= \bm\sigma \bm{\tilde{E_t}}$ with the conductivity tensor 
\begin{equation}
\sigma =
\begin{bmatrix}
\sigma_{xx} & \sigma_H \\
-\sigma_H & \sigma_{xx}
\end{bmatrix}
.
\end{equation}
$\sigma_H$ is the Hall optical conductivity and $\sigma_{xx}$ is the longitudinal conductivity. 
With the dispersion relation $\omega = cq$ we get eventually from Eqs. (\ref{maxwell1a}) and 
(\ref{bc1}) a relation between the transmitted and the reflected electric field:
\begin{equation} 
\label{maxwell2}
\bm{\tilde{E_r}}= -\frac{1}{\sigma_0}\bm\sigma \bm{\tilde{E_t}}
\end{equation} 
with $\sigma_0=2\epsilon_0c$, where we have used $\mu_0=1/\epsilon_0 c^2$.
Moreover, we get from $\bm{\tilde{E_r}}=\bm{\tilde{E_t}}-\bm{\tilde{E_i}}$ the relation
\begin{equation}
\label{maxwell2a}
\bm{\tilde{E_t}} =  \left(
\bf{1} + \frac{1}{\sigma_0}\bm\sigma\right)^{-1}\bm{\tilde{E_i}}
.
\end{equation}
These two relations will be used subsequently to calculate the transmission, reflectance and absorption
for the 2D sheet, where the latter is characterized by its conductivity tensor $\bm\sigma$. 

%

\begin{figure}[t]
\begin{centering}
\includegraphics[width=10cm]{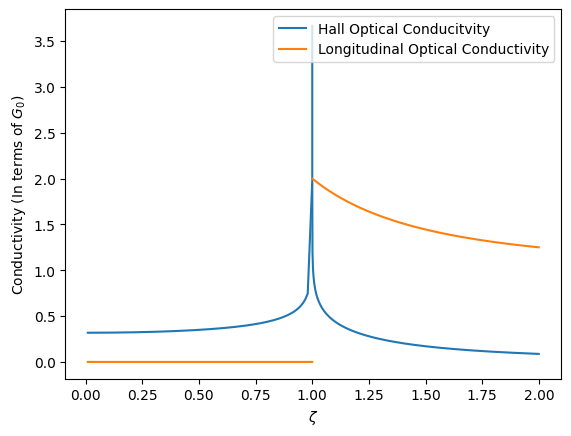}
\caption{The optical conductivities of a 2D material in units of $G_0$ with a
band gap $2m$ in the presence of the QAH effect as a function of the scaled frequency $\zeta=\hbar\omega/2m$.
}
\label{fig:1}
\end{centering}
\end{figure}

\section{Quantum transport}
\label{sect:q-transport}

Optical properties of a material are determined by the electronic transport. In the present case of a 2D
material, the optical properties are expressed by the linear response to an in-plane electric field, which oscillates in time with frequency
 $\omega$. The conductivity tensor consists of 
 the longitudinal conductivity $\sigma_{xx}$ and the Hall conductivity $\sigma_H$.
Within the Kubo formalism we obtain in the absence of thermal fluctuations~\cite{Hill_Sinner_Ziegler,nualpijit24}
\begin{equation} 
\label{hall_cond}
\sigma_H = \frac{e^2}{4h\zeta}\log{\left| \frac{1+\zeta}{1 - \zeta} \right|}
,\ \ \ \zeta=\hbar\omega/2m
\end{equation} 
and
\begin{equation}
\label{cond}
\sigma_{xx} = \frac{\pi e^2}{2h}\left ( 1+\zeta^{-2} \right )\Theta(\zeta - 1),
\end{equation}
where $e$ is the electron charge and $h$ is the Planck constant. $\sigma_{xx}$ vanishes if the
energy of the electric field $\hbar\omega$ is less than the gap energy.
Both conductivities are subject to quantum fluctuations, reflected by the universal prefactor 
$G_0=\pi e^2/2h$~\cite{nair08} with the value $h/e^2\approx 26$k$\Omega$. 
We note that $\sigma_{xx}$ decreases only by a factor of 1/2 over the entire range of $\zeta>1$
In contrast, the
classical conductivity, which is obtained from the Boltzmann theory, depends on non-universal
material properties. Comparing this result with the conventional Drude approach~\cite{ashcroft},
the classical parameter $n\tau/m$ ($n$ is the electron density, $\tau$ is the scattering time and $m$ is 
the electron mass) in the latter is replaced by the universal parameter $\pi/2h$ in quantum transport.
Moreover, the only physical parameter is the 
ratio $\zeta=\hbar\omega/2m$. Thus, the relevant energy scale of the 2D material is 
given by the gap $2m$, which is in typical
materials of the order of 1eV. The temperature-dependent conductivity depends on $\hbar\omega/k_BT$,
which does not change significantly if $\hbar\omega/k_BT>10$~\cite{ziegler07}. 
Therefore, thermal fluctuations can be neglected if this condition is satisfied.
With these expressions we anticipate that the optical properties depend on $\zeta$ with the singular
point $\zeta=1$.

The discontinuity of $\sigma_{xx}$ is the result of the photo-electric effect, through which excitation of the 
electron-hole pairs across the band gap in 2D sheet are possible: 
For a band gap $2m$ an incoming electromagnetic radiation with frequency 
$\hbar\omega < 2m$ cannot be absorbed to excite electrons from the valence band 
to the conduction band, meaning that the conduction band will be left empty. 
As a result, the material will have zero longitudinal optical
conductivity in this domain, as seen in Fig. \ref{fig:1}. However, once $\hbar\omega$
exceeds the gap energy $2m$, electron-hole pairs can be created across the band gap
due to the photo-electric effect. This allows for a non-zero longitudinal optical conductivity, 
leading to electron-hole pair excitations across the band gap.
The nature of the Hall conductivity is different, since it does not rely on the creation of electrons in the
conduction band in  the bulk of the 2D sheet but is the consequence of special correlations, as we can see
from the Kubo approach~\cite{Hill_Sinner_Ziegler}. Therefore, it is nonzero even when the energy quantum
of the electric field $\hbar\omega$ is less than the gap or even zero. This is reflected by behavior in 
Eq. (\ref{hall_cond}), which is also sensitive to the creation of electron-hole pairs.

\section{Optical properties: transmission, reflection and absorption} 
\label{sect:o-properties}

The transmission $T$ and reflection $R$ coefficients are calculated from the intensities of the transmitted
and the reflected electric fields relative to the intensity of the incoming electric field. They are written 
for $\tilde{E}_{i,t,r}^x=\tilde{E}_{i,t,r}\cos\phi$ and $\tilde{E}_{i,t,r}^y=\tilde{E}_{i,t,r}\sin\phi$ as
\begin{equation} 
\label{trans0}
T \equiv \frac{I_t}{I_i} = \frac{|\tilde{E_t}^x|^2 + |\tilde{E_t}^y|^2}{|\tilde{E_i}^x|^2 + |\tilde{E_i}^y|^2} 
= \frac{\tilde{E_{t}}^2}{\tilde{E_{i}}^2}
,
\end{equation} 
\begin{equation} 
\label{refl0}
R \equiv \frac{I_r}{I_i} = \frac{|\tilde{E_r}^x|^2 + |\tilde{E_r}^y|^2}{|\tilde{E_i}^x|^2 + |\tilde{E_i}^y|^2} 
=\frac{\tilde{E_{r}}^2}{\tilde{E_{i}}^2} 
\end{equation}
and the absorption coefficient $A=1-T-R$.
As a consequence of the assumed isotropy of the 2D sheet, there is no $\phi$ dependence.
For anisotropic materials, such as graphene-like materials under uniaxial strain, we would also get
a $\phi$ dependence in the optical properties~\cite{Nualpijit_Sinner_Ziegler,nualpijit26}.

\begin{figure}[t]
\begin{centering}
\includegraphics[width=10cm]{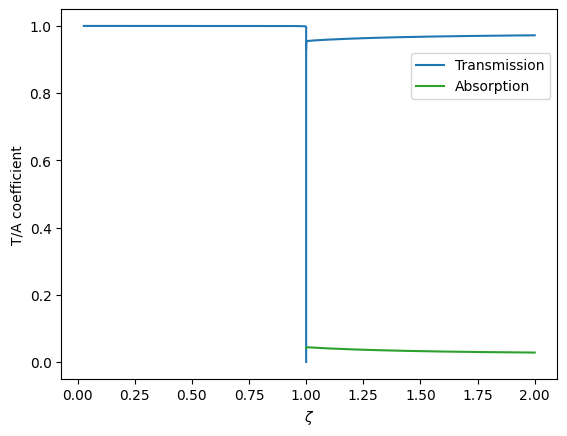}
\caption{Transmission and absorption coefficients of a 2D material with a
band gap $2m$ in the presence of the QAH effect and an incident electromagnetic wave 
with the scaled frequency $\zeta=\hbar\omega/2m$.
}
\label{fig:2}
\end{centering}
\end{figure}

\begin{figure}[t]
\begin{centering}
\includegraphics[width=10cm]{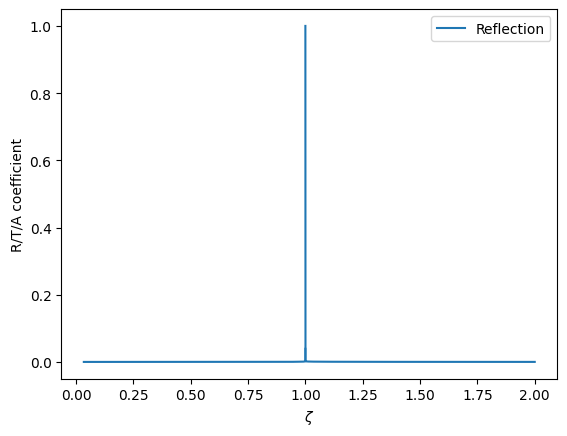}
\caption{Singular behavior of the reflection coefficient of a 2D material with a band gap $2m$ in the presence of the 
QAH effect.
}
\label{fig:3}
\end{centering}
\end{figure}

In order to calculate the coefficients, we use Eqs. (\ref{maxwell2}), (\ref{maxwell2a}) and define the 
constant
\begin{equation}
\gamma = \det(
\bm{1} + \bm{\sigma}/\sigma_0) 
= (1+s_{xx})^2 +s_H^2
\end{equation}
with $s_{xx}=\sigma_{xx}/\sigma_0$ and $s_H=\sigma_{H}/\sigma_0$ to get
\begin{equation}\tilde{E_{t}} \begin{bmatrix}
\cos\phi \\
\sin\phi
\end{bmatrix} = \frac{\tilde{E_i}}{\gamma} 
\begin{bmatrix} \left ( s_{xx}+1 \right )\cos\phi - s_H\sin\phi \\ s_H\cos\phi + \left ( s_{xx}+1 \right )\sin\phi 
\end{bmatrix}
\end{equation}
and
\begin{equation}
\tilde{E_{r}} \begin{bmatrix}
\cos\phi \\
\sin\phi
\end{bmatrix} = \frac{\tilde{E_i}}{\gamma} 
\begin{bmatrix} -(s_{xx}^2+s_H^2 + s_{xx})\cos\phi - s_H\sin\phi \\ 
s_H\cos\phi  -(s_{xx}^2+s_H^2 - s_{xx} )\sin\phi 
\end{bmatrix}.
\end{equation}
Then, using the definitions in Eqs. (\ref{trans0}), (\ref{refl0})
we obtain for the transmission and reflection coefficients:
\begin{equation} 
\label{trans2}
T = \frac{1}{(1+s_{xx})^2+s_H^2}
,\ \ \
R = \frac{s_{xx}^2+s_H^2}{(1+s_{xx})^2+s_H^2}.
\end{equation} 
Finally, the optical absorption coefficient reads
\begin{equation}
A 
= \frac{2s_{xx}}{(1+s_{xx})^2+s_H^2}
,
\end{equation} 
which vanishes with a vanishing $\sigma_{xx}$.
Now we can obtain $T$, $R$ and $A$ as functions of the frequency-dependent parameter
$\zeta$ by calculating the optical conductivities defined by 
Eqs. (\ref{hall_cond}) and (\ref{cond}). With $\sigma_0=2\epsilon_0c$
we can write $G_0/\sigma_0=\pi\alpha/2$ with $\alpha=e^2/2\epsilon_0 hc\approx 1/137$, 
also known as the fine-structure constant~\cite{sommerfeld1916}.
Thus, both dimensionless conductivities are proportional to the fine-structure constant, where
the dimensionless Hall conductivity reads
\begin{equation}
s_H = \frac{\alpha}{4\zeta}\log{\left| \frac{1+\zeta}{1 - \zeta} \right|},
\end{equation} 
and the dimensionless longitudinal conductivity is 
\begin{equation}
\label{cond2}
s_{xx}= \frac{\pi\alpha}{2}\left ( 1+\zeta^{-2} \right )\Theta(\zeta - 1)
.
\end{equation}
These general results can be reduced to some special cases. First, we consider $s_{xx}=0$, where 
there is no absorption and the reflection is proportional to the transmission:
$T=1/(1+s_H^2)$, $R=s_H^2T$ and $A=2s_HT$.
Second, for $s_H=0$ the absorption has a maximum at $\zeta=1$ and decays slowly for $\zeta>1$.
On the other hand, for $s_H>0$ the absorption vanishes at $\zeta=1$ and its 
maximum is just above $\zeta=1$ (cf. Fig. \ref{fig:2}).
Finally, for a vanishing gap (i.e., in the limit $\zeta\to\infty$) $s_H$ vanishes and the coefficients read
\begin{equation}
T=\frac{1}{(1+\pi\alpha/2)^2}
, \ \
R=\frac{(\pi\alpha/2)^2}{(1+\pi\alpha/2)^2}
,\ \ 
A=\frac{\pi\alpha}{(1+\pi\alpha/2)^2}
\end{equation} 
with 97.7\% transmission, 2.3\% absorption and almost no reflection
was experimentally confirmed for graphene~\cite{kuzmenko08,nair08}. 
This is a remarkable result, since it is constant in the frequency of the external field. Measuring $T$
provides a useful tool for determining the fine-structure constant experimentally.

\section{Discussion and conclusion}
\label{sect:discussion}

As illustrated in Fig. \ref{fig:2}, 
there is no absorption for $\zeta<1$ due to $s_{xx}=0$, while the reflection almost vanishes for all $\zeta\ne 1$
except for the singularity at $\zeta=1$, where it jumps to $R=1$ (cf. Fig. \ref{fig:3}).
For $\zeta>1$ absorption appears due to $s_{xx}>0$ and decreases monotonically for larger values of the
scaled frequency $\zeta$.  At the same time, the transmission is close to 100\% for $\zeta<1$, jumps to $T=0$
at $\zeta=1$ and increases monotonically for $\zeta>1$, since the absorption decays slowly for higher 
frequencies. The absorption always exceeds the reflection but never the transmission, as illustrated in Fig. \ref{fig:2}.

Turning off the Hall conductivity $s_H$ does not change the optical behavior away from a very narrow region 
around $\zeta=1$. 
Thus, for $\zeta<1$ the 2D sheet is completely transparent, since $s_{xx}=0$ in this case,
representing an ideal insulator in terms of quantum transport. On the other hand, for $\zeta>1$ all three 
coefficients jump to some nonzero value, reflecting that the 2D sheet becomes metallic. The transmission is above
97\%, the absorption is less than 3\% and the reflection is below 1\%.
With increasing frequency the transmission increases, while the reflection and the absorption decrease.

In conclusion, we found a remarkably simple behavior for the optical properties of a gapped
2D sheet. As a function of the frequency of the electric field, the properties
are characterized by 100\% transmission for $\omega<2m/\hbar$ and 97\% transmission and 3\% absorption for
$\omega>2m/\hbar$. This could be useful for a number of applications, in which the gap energy must be accurately
determined by an optical probe.

\end{document}